# Amphibious Transport of Fluids and Solids by Soft Magnetic Carpets

Ahmet F. Demirörs[1]*, Sümeyye Aykut[1], Sophia Ganzeboom[1], Yuki Meier[1], Robert Hardeman[2], Joost de Graaf[2], Arnold J.T.M. Mathijssen[3], Erik Poloni[1], Julia A. Carpenter[1], Caner Ünlü[4], and Daniel Zenhäusern[5]

[1]Complex Materials, Department of Materials, ETH Zurich, 8093 Zurich, Switzerland

[2]Institute for Theoretical Physics, Center for Extreme Matter and Emergent Phenomena, Utrecht University, Princetonplein 5, 3584 CC Utrecht, The Netherlands

[3]Department of Physics and Astronomy, University of Pennsylvania, Philadelphia, PA 19104, United States

[4]Department of Chemistry, Istanbul Technical University, Istanbul, Turkey

[5]Institut für Solartechnik SPF, HSR University of Applied Sciences Rapperswil, Rapperswil, Switzerland

*Corresponding author. Email: ahmet.demiroers@mat.ethz.ch

**Abstract:**

One of the major challenges in modern robotics is controlling micromanipulation by active and adaptive materials. In the respiratory system, such actuation enables pathogen clearance by means of motile cilia. While various types of artificial cilia have been engineered recently, they often involve complex manufacturing protocols and focus on transporting liquids only. Here, we create soft magnetic carpets via an easy self-assembly route based on the Rosensweig instability. These carpets can transport liquids but also solid objects that are larger and heavier than the artificial cilia, using a crowd-surfing effect. This amphibious transportation is locally and reconfigurably tuneable by simple micromagnets or advanced programmable magnetic fields with a high degree of spatial resolution. We identify and model two surprising cargo reversal effects due to collective ciliary motion and non-trivial elastohydrodynamics. While our active carpets are generally applicable to integrated control systems for transport, mixing and sorting, these effects could also be exploited for microfluidic viscosimetry and elastometry.

**Introduction**

Transport of solids and fluids is key to maintaining continuous processes in the microscopic and macroscopic world, *e.g.*, mucociliary transport for the clearance of pathogens out of the respiratory system[1], the locomotion of reproductive cells by motile flagella[2]; and the transport of goods in industry and everyday life[3]. Transport generated by soft actuators is of interest in many areas ranging from soft robotics[4] to drug delivery[5] and microfluidics[6]. In robotics, transport of solid (and fragile) objects has been achieved using soft actuators[7–10]. Recent demonstrations thereof include pneumatic designs[11,12]. However, often these soft actuators have to be driven by complicated algorithms[12,13] to achieve simple rotational or bending motions. In addition, pneumatically activated systems usually lack autonomy, due to being tethered to a pump[14]. Fully autonomous soft actuators exist, although these can suffer from fuel depletion upon long-lasting activation[9,15]. These considerations limit the general use of pneumatic actuators. In contrast, systems driven by external fields generally perform better in terms of durable activity[8,16] and autonomy[17–19]. This has led to field-driven soft actuators, especially ones driven by magnetic fields, receiving significant attention recently[20–24], also in view of the range of forces that can be applied to transportable cargos using such systems[25].

Fluid transport and mixing by means of whip-like organelles, cilia, is an efficient soft actuation strategy found throughout nature: from microorganisms[26] to mammalian airways[1,27,28] to brain ventricles[29]. In the microfluidic regime, viscosity dominates inertia, rendering conventional pumping of fluids highly inefficient. The efficiency of the ciliated fluid transport has motivated scientist to mimic this strategy, primarily to replicate the biological functionality[30–35]. Additionally, biomimetic cilia can be driven by external fields with relative ease[30–32,36,37]. The state of the art in this regard is the work on solid-cargo transport using magnetic soft cilia[36,38]. However, this study focused on micron-sized colloidal objects suspended in a liquid; dry cargo transport remains unexplored. Another recent study showed the transport and manipulation of water droplets using magnetically responsive board-like structures[38]. Yet, the fabrication of these boards is difficult to scale up and was limited to unidirectional transport. Many realizations of magnetically actuated artificial cilia are useful for specialized functionalities, but are often expensive or involve time intensive fabrication techniques, e.g., lithography[36], 3D printing[30], femtosecond laser writing[38], and manual assembly[32]. This can significantly limit the scalability of these approaches.

Here, we create arrays of soft responsive pillars, which we collectively refer to as a soft magnetic carpet (SMC). Their manufacture is based on a hard ferromagnetic variant of the facile fabrication method outlined by Lu *et al.*[39] and Timonen *et al.*[40], which involves an external magnetic field and a resin doped with magnetic particles (see Methods for details). Pillars emerge with the application of the field, due to the Rosensweig instability[41], and possess a permanent magnetic dipole moment after fabrication. A major advantage of this approach is its scalability. The permanent magnetic moment can be used to actuate the SMCs using a (patterned) external magnetic field. The use of silicone resins enables the 'amphibious' transport of both liquids and solids, such as millimetric droplets and large objects, as well as the generation of fluid flows when the carpet is fully submerged. In both the dry and wet (submerged) states, there are interesting reversals of transport that we characterize and model theoretically. Owing to their simple design, scalable fabrication scheme, and the level of achievable fine control over transport, sorting, and mixing, we foresee a myriad of potential applications of these SMCs in industrial and academic settings.

## Results

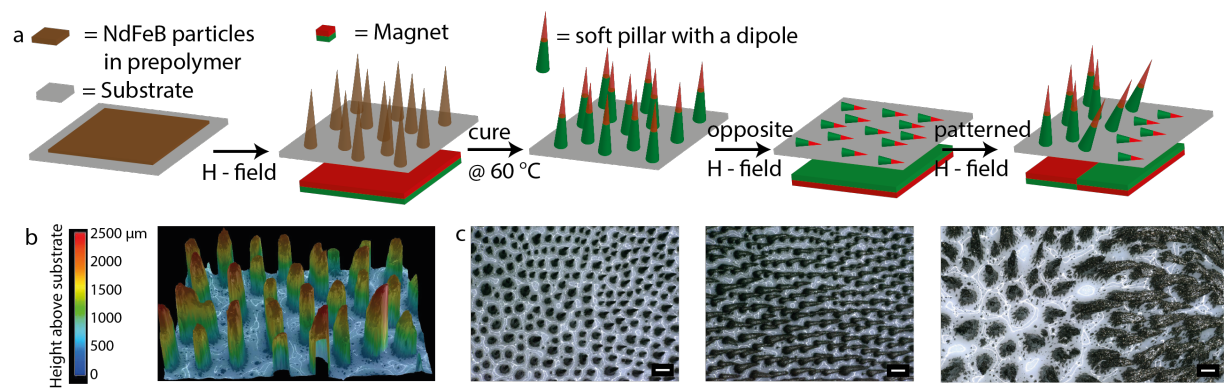

**Figure 1: Fabrication of soft magnetic carpets.** a) Sketches of the fabrication steps of our SMCs, depicting how the magnetic prepolymer forms pillar-like structures that become stable after being cured at 60 °C. The pillars carry a permanent magnetic dipole moment and therefore respond to external magnetic fields. b) A reconstructed 3D reflection microscopy scan of the SMC; no magnetic field is applied here. c) Reflection microscopy images of the pillars sketched in (a), without an external magnetic field, with an oppositely directed field (compared to the formation), and with a half similarly and half oppositely directed field, respectively. Scale bars are 200 μm.

We fabricated an array of soft pillars with a permanent magnetization following a modification of the procedure outlined in[39,40], see Methods for a full description. In brief, we spread a mixture of neodymium iron boron (NdFeB) particles (Magnequench MQFP-B+, D50 $\approx$ 25 μm) and a soft silicone matrix (Ecoflex 00-20) on a substrate, see Figure 1a. To this layer, we applied a homogeneous magnetic field during the course of the polymerization. This caused the spontaneous formation of pillars due to a Rosensweig-type instability[40]; Figure 1b shows a micrograph of the final shape. The size and density of the pillar array can be tuned readily by changing the applied field strength, the magnetic particle content, and the total amount of the magnetic mixture (see Supplementary Table S1 for details). Once polymerized fully, the pillars respond to external magnetic fields, as they possess a permanent magnetic dipole moment imbued by the homogeneous external field during fabrication through the alignment of the moments of the individual NdFeB particles. When an external field is applied parallel to the pillars' dipole moment, the pillars stretch in the field gradient. When the magnetic field is oppositely directed, it bends the pillars onto the substrate surface, see Figure 1c. We found this response to be reversible, due to the pillars' softness, and easy to tune on a millimetric scale by designing a magnetic actuation landscape, similar to the recent work[35]. These features can be used to induce pillar motions that result in the transport of fluids and solids; here, through the application of a traveling oscillatory magnetic field. Such magnetic actuation allows one to perform work, as described next.

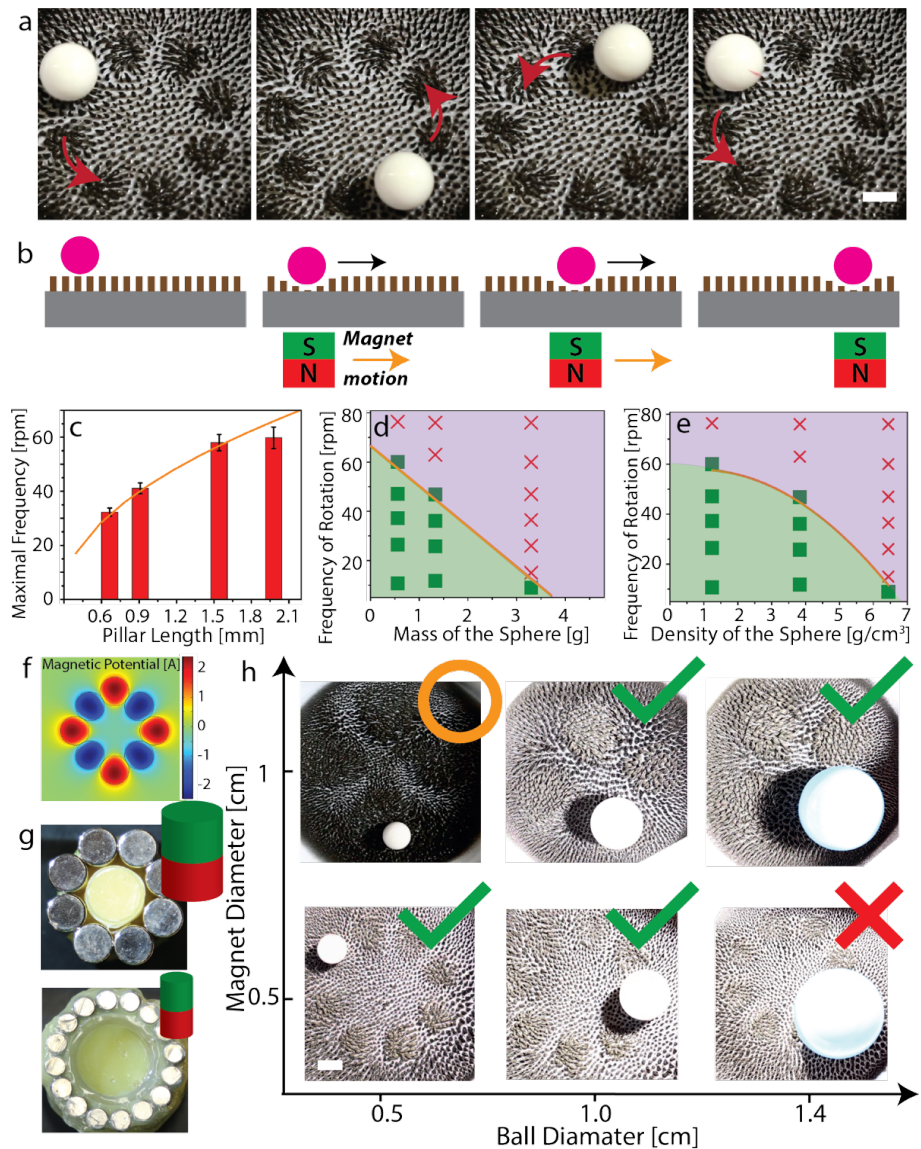

**Figure 2: Micromanipulation of solid objects on dry soft magnetic carpets.** a) Snapshots showing circular motion of a spherical object by the rotation of a set of alternately oriented rod-shaped magnets (magnetic track) placed underneath the substrate covered by the SMC. b) Sketches indicating the way the SMC behaves under the application of a localized magnetic field and how this induces transport of a sphere supported by the SMC. c) Pillar-length dependence of the magnetic track's maximum angular frequency $\omega$ for which sphere transport occurs. d) State map of the sphere's ability to follow the motion of the magnetic track (green squares = 90% co-rotation, red crosses < 50% co-rotation), plotted as a function of $\omega$ versus the sphere's mass. e) State map of $\omega$ versus the sphere's mass density. c-e) The orange curves show our theoretical prediction for the co-rotation crossover. f-h) Effect of the magnet size in relation to the sphere diameter on the transport performance. f) The gradient map provides the finite-element calculated magnetic potential of the 8-magnet track at the surface of the substrate

showing that four natural wells will form. g) Photographs of the magnetic tracks. h) Photographs of the associated carpet patterns, with four and eight magnetic wells, respectively, and the spheres used. Green checks indicate co-rotation (less than 10% well hopping), the red crosses indicate more than 80% well hopping, and the orange circle a roughly 50% rate of well hopping. All scale bars indicate 5 mm.

Our SMCs can be used to manipulate solid non-magnetic objects in space. To demonstrate this, we arranged cylindrical magnets in a ring shape (magnetic track), alternating their orientation. This magnetic track was rotated underneath the substrate covered by the magnetic pillars, on which a non-magnetic millimetric sphere rested. The field induced by our magnetic track formed a ring-shaped array of depressions (wells) on the surface of the carpet as shown in Figure 2a-b. These wells were formed above places where the magnetic field was directed opposite to the orientation of the homogeneous field applied during fabrication, *i.e.*, the intrinsic magnetization direction of the pillars. As such, there are half the number of depressions than there are magnets in the track, see Figure 2f,g. The wells were made mobile by rotating the track. They are a gravitationally favorable location for a solid object, as sketched in Figure 2b. Hence, the motion the track causes the spheres to co-rotate, provided they do not escape the well, see Supplementary Movie S1. Transport by this mechanism depends on several factors of which the most important are: the length of the pillars; the mass, density, and the size of the sphere; and the size and magnetic strength of the magnets. First, we analyzed the transport capacity of our SMCs as a function of the rotation frequency of the magnetic track, as well as pillar length. In all cases we used a 9.5 mm sized sphere with a mass of 0.56 g. We plot the maximal angular frequency at which the sphere remains confined to its well in Figure 2c. The shorter the pillars (average height: 670 µm, 910 µm, 1.5 mm, and 2.0 mm, respectively), the less effective the wells were in retaining the sphere. Secondly, we used similarly sized spheres with different mass densities to explore their effect on the SMC's transport ability (see Methods for details). We mapped successful (green points, less than 10% escapes) and unsuccessful rotations (red points, an escape rate exceeding 50%) in mass-angular frequency (Figure 2d) and density-angular frequency (Figure 2e) diagrams. Our SMCs were found to transport the lower mass and lower density spheres more readily. Finally, we have checked the effect of the ratio of magnet size to cargo size, *i.e.*, the effect of the width of the well. For these experiments, we kept the pillar height constant at 2 mm. Figure 2f shows that matching the size of the magnet to the ball leads to better transport performance, compared to strong mismatches. A relatively small well cannot provide sufficient space to capture a large sphere. Conversely, when the well is much larger than the sphere, the sphere has significant freedom to move and can gain enough momentum in the well to jump over the edge.

We can understand the above result using a minimal model for the maximal work that can be harvested from our SMC system. In our model, we consider the maximum transport capacity to depend on three factors: (i) pillar stiffness, (ii) pillar susceptibility to the magnetic field, (iii) the pillar susceptibility to pressure. The former and latter two determine the height difference between the straight and bent pillars. We identify the potential energy difference between the sphere resting on the two pillar conformations as the main contributor to localizing an object (sphere) to a well, see Supplementary Figure S1a. When the sphere's kinetic energy exceeds this potential energy gain, the sphere is expected to escape the well. Therefore, the magnetic track's maximum angular frequency $\omega$ at which corotation breaks down, is found by equating the sphere's kinetic ($K$) and potential energy ($P$), $i.e.$, $K = P$. A sphere rotating with a constant frequency along the circular path imposed by the track has a kinetic energy $K = \frac{1}{2} mR^2\omega^2$, where $m$ is the mass of the sphere and $R$ is the track radius. The potential-energy gain due to the magnetitic deflection of the pillars is given by $P_m = mg\Delta h_m$, where $g$ is the acceleration due to gravity, $\Delta h_m$ is the height difference between the center of mass of the sphere resting on straight pillars and at the bottom of the well, see Supplementary Figure S1b. A sphere is expected to make the deeper by its mass resting on the pillars. This gives rise to an additional potential energy difference $P_w = mg\Delta h_w$, where $\Delta h_w$ is the mass-induced height difference. We assume that the pillars act as a Hookian spring, $F = kx$, with $x$ the deviation from equilibrium, and spring constant $k$ representing the stiffness of the silicon material (Ecoflex 00-200), see Supplementary Figure S1c. Writing the magnetic force as $F_m$, we obtain $\frac{1}{2}mR^2\omega^2 = \frac{mg}{k}(F_m - mg)$. Solving this equation for $\omega$ results in the mass (and density dependences) shown by the orange curves in Figure 2c-e. In all cases, the shape agrees well with the experimental findings.

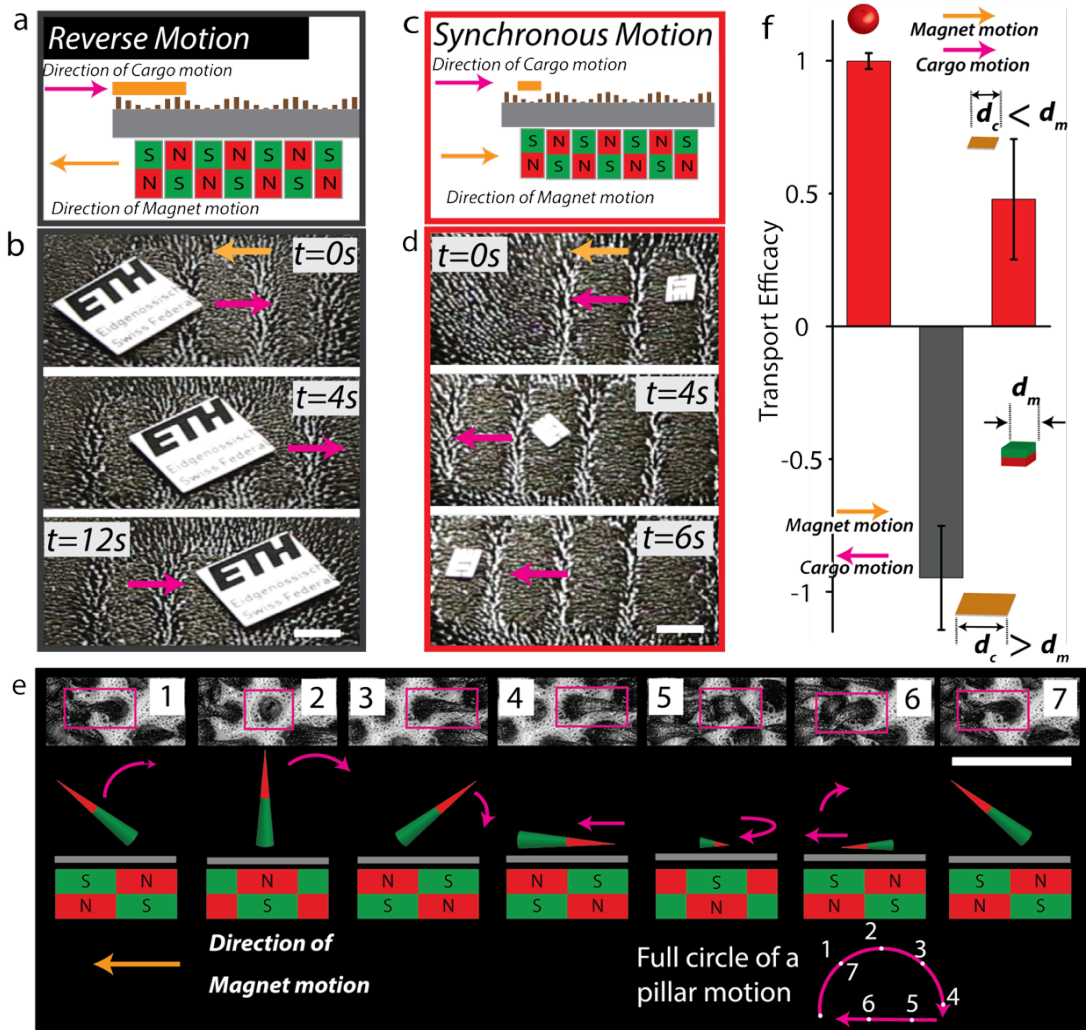

**Figure 3**: **Size-dependent transport on dry SMCs**. a) Sketch of a SMC carrying the board-shaped cargo in the opposite direction to the motion of the driving magnets. This demonstration is for an object with a contact-dimension larger than that of a single magnet in the array. b) Time lapses showing the motion of a large board-shaped cargo on a SMC. c) Sketch of a SMC carrying a small board-shaped cargo along the same direction as the motion of the driving magnets. d) Time lapses showing the motion of a small board-shaped cargo on a SMC. e) Time lapse of microscopy images showing a full cycle of pillar conformations during the motion of the magnet track toward the left. The top-down images are accompanied by sketches indicating the position of the magnets in the track and the response of a single pillar. d) Transport efficacy (object motion relative to that of the magnetic track) for the spherical cargo and two sizes of board-shaped cargos. Scale bars indicate 5 mm in (b,d) and 500 μm in (e).

Our SMCs can transport a wide variety of shapes. The transport of a solid cargo depends sensitively on the shape of the cargo and in particular on the dimension of the object with respect to the natural wavelength of the wells in the SMC, see Figure 3a-d. If the dimension of the contact is smaller than the magnetic-well size, the cargo co-moves with the magnets, see Figure 3c,d, Supplementary Figure S2 and Movie S2. However, intriguingly, for anisotropic shapes of sufficient size, like a cylinder or a board, cargo locomotion is opposite to that of the magnet track, see Figure 3a,b (also Supplementary Movie S3 and S4). To understand this behavior, we considered the motion of a single pillar as a function of the motion of the magnets, similar to what was done in the recent study[31]. The time lapse in Figure 3e shows the associated sequence of pillar conformations. Here, microscopy images together with cross-sectional sketches demonstrate that the pillar motion is highly anisotropic and covers a half circle, see inset Figure 3e. In a 'forward stroke' (power stroke), steps 1 to 4, the pillar stretches and moves in the vertical direction, conversely in the 'backward stroke' (return stroke), steps 5 to 7, the pillar shrinks and stays close to the substrate. We use these naming conventions to make the connection to the stroke pattern in biological cilia. Note that despite the naming, the forward stroke is, in fact, in the opposite direction to the motion of the magnet. Therefore, a cargo smaller than a single magnet unit (*i.e.*, smaller than the magnetic well) travels in the same direction as the magnet. A cargo larger than a single magnet comprising the track does not fit inside the well and instead predominantly experiences the motion of the stretched pillars, which perform a power stroke, as depicted in steps 1-4. This causes such a large cargo to travel in opposite direction to the magnetic track in a manner that is reminiscent of crowd surfing. Transport efficacies for board-shaped and spherical cargos are given in Figure 3f. It is clearly observed here that a board-like cargo that is smaller than the magnets, moves in the same direction as the magnets. However, a cargo that is larger moves in the opposite direction (there denoted as having a negative efficacy). The dual nature of the motion as a function of pillar contact can be used to separate out particles with different shapes and sizes. For instance, a ring-shaped object can be sorted from spherical objects using our soft-carpet setup, see Supplementary Figure S3 and Supplementary Movie S5. Additional cargo shapes and associated transport efficiencies are also discussed in Supplementary Figure S3.

Solid cargo transport can also be performed over a tilted surface. We tilted the substrate at a 20° angle to the surface, in which case the soft carpet transported an alumina cylinder upward, as shown in Supplementary Figure 2 and Supplementary Movie S6. The maximum slope over which transport may be achieved depends on the length of the pillars and its comparison to the size of the cargo. Generally, we expect that an increase of the ratio of pillar length to cargo size will result in improved performance on steeper slopes.

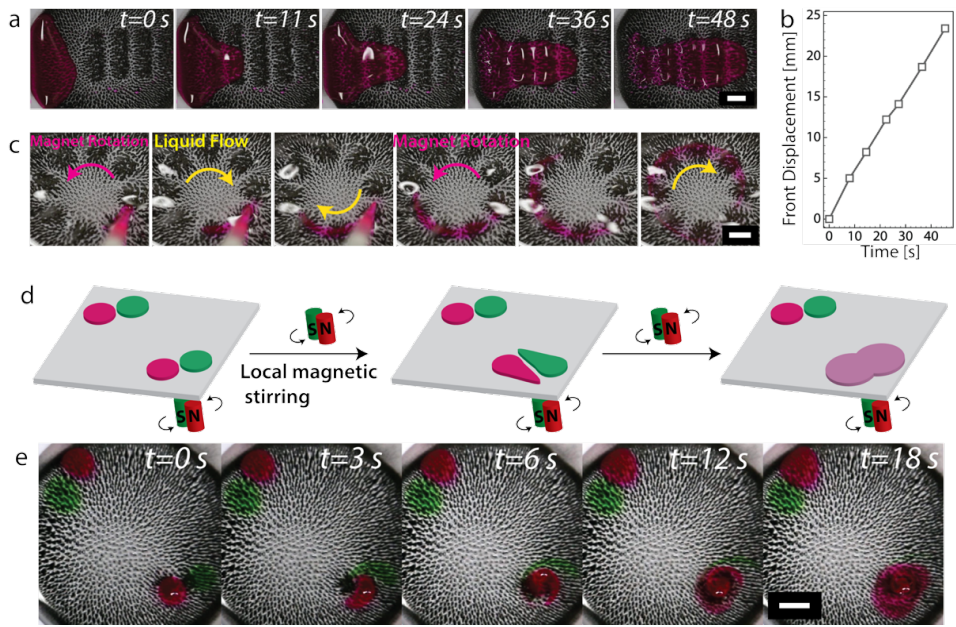

**Figure 4: Droplet transport, fluid flow, and viscous mixing on SMCs.** a) Photographic time lapse taken of a glycerol droplet (labelled with rhodamine) driven over a dry SMC by a linear track of magnets. The droplet moves in the opposite direction to the magnetic track. b) Plot of the displacement of the glycerol droplet's front with time. c) Continuous flow of glycerol (fully submerged SMC) driven by a circular magnetic track. The motion is in the direction opposite to that of the track. The direction of flow was visualized by adding a droplet of rhodamine labelled glycerol. d) Sketch demonstrating how soft carpets enable spatial control over local fluid mixing. Glycerol droplets labelled with different dyes (rhodamine and green food color) are placed at opposite corners of the milli-fluidic chip. Active mixing takes place in one corner while the fluid in the opposite corner remains effectively unperturbed. e) Photographic time lapse taken of a local mixing experiment that demonstrates the significant enhancement of SCM-based mixing compared to that induced by diffusion in glycerol. Scale bars are 5 mm.

In addition to solid transport, our (otherwise dry) SMCs can transport and interact with liquid droplets. To demonstrate this, we placed a small amount of glycerol on one of our SMCs and labelled the fluid with rhodamine dye to visualize the transport. The time lapse shown in Figure 4a demonstrates that the glycerol droplet moves in opposite direction to the motion of the magnetic-field wave. Figure 4b quantifies the motion of the droplet front; the liquid moved over 23 mm in less than a minute, see Supplementary Movie S7. To induce a continuous flow, we implemented a ring-shaped magnetic track and rotated this underneath a SMC fully submerged by glycerol. Figure 4c shows that this resulted in a liquid counter-flow opposite to the direction of the track rotation, also see Supplementary Movie S8. This is reminiscent of antiplectic metachronal motion in biological cilia[31,32], but as we will see shortly not necessarily indicative of metachronicity. Note that the direction of the movement, along or opposite to the track motion, depends on the liquid viscosity as explained in the subsequent sections.

Diffusive mixing of solutes in viscous liquids such as glycerol is generally inefficient due to the laminarity of the flow at low Reynolds numbers, *i.e.*, viscous dissipation dominates inertia. This is a limiting factor in achieving conventional mixing in microfluidic chips, capillaries, and microfluidic devices[6]. The non-reciprocal motion of biological and biomimetic cilia overcome this obstacle. Here, we demonstrate that our SMC are not only capable of achieving mixing at a microfluidic scale, but that this can be done with precise spatial control. We placed two droplets of glycerol with two distinct dye labels (rhodamine and green food color) in opposite corners of a SMC submerged in glycerol. A local stirring flow was induced at one corner using a pair of magnets. Over the course of 20 to 30 seconds, the droplets could be fully mixed by actuating the magnets, while the situation in the opposite corner remained relatively unperturbed, see Figure 4d,e and Supplementary Movie S9.

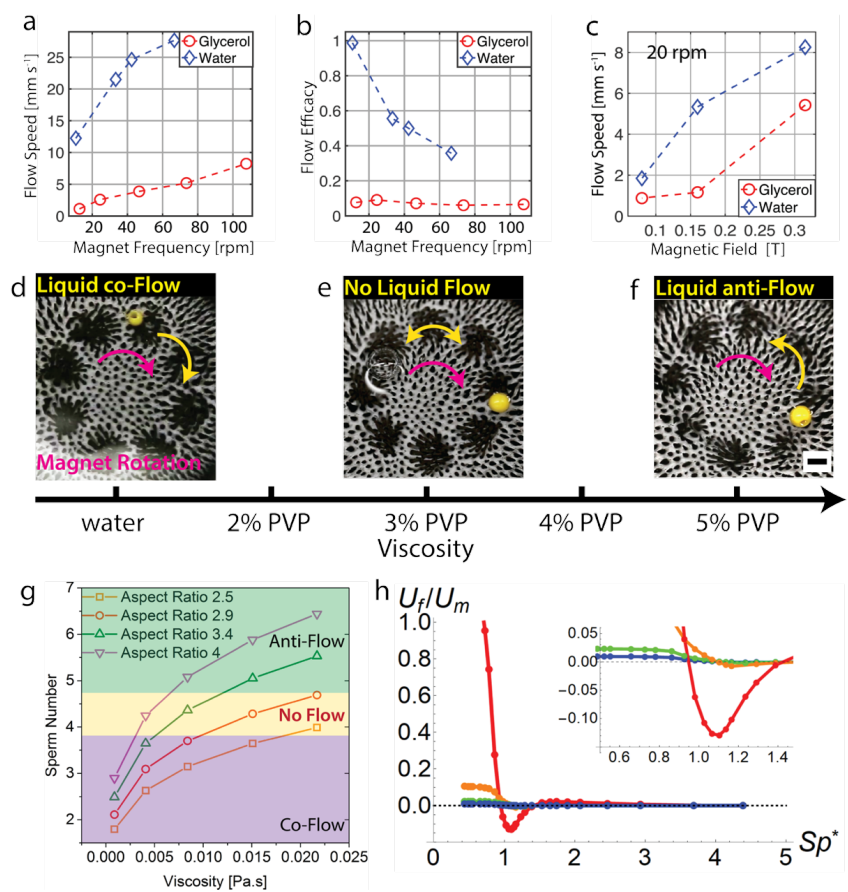

**Figure 5. Non-trivial fluid transport properties.** a) Dependence of the flow speed on the angular frequency of the magnetic track. b) Dependence of the transport efficacy on the angular frequency. c) Dependence of the flow speed to the magnetic field strength at affixed 20 rpm magnet frequency. d-f) Snapshots showing three instances of a sphere resting on a fluid layer that is actuated by a submerged SMC. The arrows show the direction of the magnetic track (purple) and the resultant liquid flow (yellow). The viscosity of the fluid increases from left to right leading to distinct flow regimes: d) DI water (comoving flow); e) 3% aqueous PVP solution (no net flow; stall-flow); f) 5% aqueous PVP solution (oppositely directed flow). g) Plot of the Sperm number, see main text for definition, for different aspect ratio magnetic pillars against the viscosity of the medium. h) Result of our bead-spring model showing the time-averaged fluid velocity of the interface $U_f$ (reduced by the speed of an effective magnet on the circular track $U_m$) as a function of the effective sperm number $Sp^*$ (where high $Sp^*$ corresponds to high viscosity and vice versa). From red to blue, the ratio of the cilium length $L$ to fluid height $H$ is $L/H$ 2/3, 1/2, 1/3, 1/4 respectively. The inset shows the resulting flow reversal in detail. Scale bars are 5 mm.

Generally, wet SMCs induce a fluid flow whose speed depends on the magnetic track's angular frequency, the strength of the applied magnetic field, and the viscosity of the submerging fluid. The fluid flow was measured by tracking the motion of a sphere supported by the liquid-air interface and found to be proportional to the angular frequency, see Figure 5a. The maximum flow speed generated was around 2.7 cm s$^{-1}$ at an angular frequency of 60 rpm for water and it was about 0.8 mm s$^{-1}$ for glycerol. The flow speed increased with increasing angular frequency, but appeared to plateau for our highest frequencies, this is reminiscent of some of the observations found in[31,35]. We also considered the flow efficacy. When the fluid makes one round for a single rotation of the magnetic track, the efficacy equals 1. Such efficiencies were only achieved at very low angular frequencies in water (around 10 rpm). The flow efficacy for glycerol was around 0.07 and remained relatively constant over the entire frequency range that we considered, see Figure 5b. Lastly, the flow speed increased by increasing the magnetic field applied to the wet SMC, see Figure 5c; we achieved this by adjusting the distance between the wet SMC and the magnetic track. We observed a flow reversal by switching from water to glycerol, see Figure 5d-f, analogous to the one reported in[35]. We will explore this phenomenon next.

In a low-viscosity medium, such as water, the flow was generally found to be in the same direction as the rotation of the magnetic track. This is reminiscent of symplectic ciliary transport in biology[42,43]. When using a sufficiently high viscosity liquid to submerge the SMC, such as glycerol, the flow was reproducibly found to be in the opposite direction. The possibility of such a reversal has been predicted for the non-reciprocal motion of a single (artificial) cilium and several (metachronally moving) cilia in various settings [44–46]. Here, we go beyond the results of Ref. [35] by examining a range of viscosities, we do so by preparing a series of aqueous PVP (360k) solutions up to 20% of PVP by weight (corresponding viscosities in Supplementary Table S2). For a SMC with a pillar aspect ratio of 3.1, the liquid flow in DI water (0% PVP) was co-moving, see Figure 5d (Supplementary Movie S10). This behavior persisted up to and including 2% PVP. For 3% PVP no net liquid flow was observed (stall flow), see Figure 5e (Supplementary Movie S11). When the viscosity was further increased (5% PVP), net fluid flow was reestablished, but it was counter to track motion, see Figure 5f (Supplementary Movie S12). We further found that by changing the aspect ratio of the pillars of the SMC, the viscosity at which stall flow occurred could be shifted: increasing the aspect ratio decreases the viscosity at which the flow behavior changes. For very high aspect ratios, the flow was observed to be oppositely directed even in DI water. We quantify this counterintuitive behavior using the Sperm number $S_p$. This is a dimensionless number characterizing the relative effect of viscous to elastic stresses on acting on a filament[47]; see the supplementary information for an estimate of this number. Figure 5g shows the effect of the aspect ratio on the flow behavior. Examining the state diagram of $S_p$ as a function of viscosity, Figure 5g, reveals that the no-flow region is found in narrow band around $S_p \approx 4$, see also Supplementary Table S2.

We will now provide arguments by which to understand this result. We first noted that the interface showed some undulation during the counter-flow in the experiment. However, a dye test proved that the time-averaged flow is indeed reversed (Figure 4c), indicating that the small oscillatory motion of the interface is not the mechanism by which the reverse transport of the bead occurs. That is, our energy-based argument for transport over dry soft carpets cannot be applied to wet soft carpets. We instead understand the observed viscosity-based transport reversal through a computational fluid dynamics (CFD) study in the spirit of Refs.[37,45]. Details of the modeling are provided in the supplementary material and Supplementary Figures S4-6. More detailed analyses of cilia motion are possible, i.e., by adapting the work of[37,45], however, this proved unnecessary for the purposes of understanding the reversal. In brief, we use a hydrodynamic Greens' function approach to capture the dynamics of the cilium, but we modify the hydrodynamic singularities to account for the finite height of the fluid layer following Ref.[48]. This is necessary to show that any mobility reversal — already observed as a function of the sperm number close to a single cilium in a fluid half space[37] — still occurs at the surface of the confining fluid layer. The extension makes our model computationally more involved, and we therefore focus on the dynamics of a single sphere (bead) only, which is located at the tip of the cilium and represents its entire motion. This choice in modeling ciliary dynamics is inspired by Ref.[49]. Our single bead is actuated to move along a closed path through an oscillatory driving that accounts for the presence of the magnets. We induced a non-linear angular spring to prevent the bead from contacting the wall and a similar longitudinal spring to model the reorientation and extension/contraction of the cilium induced by the driving. This form of driving and choice of springs proved sufficient to capture the salient features of the experiment, see Figure 5H, which shows a reversal of the flow velocity with the sperm number for various heights of the fluid layer. The reversal can be understood by the interplay of a reduction in bead mobility near the surface, the restorative spring, and the interaction with the magnets. This is evidenced by the roughly unitary value of the effective sperm number $Sp^*$ at which the primary reversal occurs. We use a slightly different choice here than in the experiment as is further clarified in the Supplementary Material. At high viscosity, which corresponds to high $Sp^*$, the cilium does not have a large deflection (as in the experiment), thus the net velocity of the interface comes purely from the asymmetry in mobility between the forward and backward strokes. The back stroke is favored here, because, while the mobility may be reduced closer to the wall, the amount of applied force is increased. At lower viscosities, i.e., lower $Sp^*$, the cilium can maximally deflect, thereby coming close to the surface. This leads to a reweighting of the forward and backward components of the stroke. The reduced mobility close to the surface leads to there being a retention time, such that it takes a sufficiently long application of the magnet to push the bead away again. This retention eliminates the most powerful part of the return stroke, thus giving rise to net forward motion of the interface just above the anchoring point. Because we have only considered a single cilium here, the fluid velocity drops rapidly with a greater fluid height. This is unlike the experiment, wherein many cilia are simultaneously actuated across the surface. Nonetheless, our single-cilium result reproduces the reversal and indicates that a metachronal pattern is not necessary for the generation of the mobility reversal in situations where the soft carpet is submerged, though we do not eliminate this possibility as a source of mobility reversal.

**Conclusions**

We have introduced soft magnetic carpets that consist of arrays of millimetric magnetic pillars that can be fabricated using a facile and scalable self-assembly route. When actuated using a suitable external magnetic field, they can transport solid objects and liquid droplets, as well as generate fluid flows for wireless microfluidic mixing. All of these can be controlled with a fine degree of spatial resolution using straightforward methods. Therefore, our system is readily advantageous to scientists with different levels of micro-robotics experience, with potential for a wide range of applications in cell manipulation, biomechanical manipulation, or microfluidic pumping and mixing. We found interesting forms of transport reversal both in the dry and wet state, which hold promise for an additional degree of control. In their dry state, this property was shown to enable the sorting and separation of solid objects depending on their size and shape. Simple and fine control over the structure of the traveling magnetic field allows our carpets to separate and sort a wide size and shape range of objects. We further showed that our carpets are able to transport objects up an incline. These features lend soft magnetic carpets the potential for the transportation and sorting of fragile objects in industrial production and assembly lines.

**Methods**

*Fabrication of the soft carpet:* We modified the method of Refs. [39,40] to fabricate our soft, magnetic pillars. First, we prepared a mixture of Ecoflex 00-20 (a Platinum Silicone rubber compound, Smooth-on Inc.), hexane (Sigma Aldrich, ACS reagent), and magnetic neodymium iron boron particles (NdFeB, Magnequench MQFP-B+, D50 = ~25 µm, 10215-088, Lot #F00492) with a weight ratio of 4:1:x, where x is set by the NdFeB particle concentration. In a typical experiment, we added 4 g of Ecoflex (2 g of each component), 1 g of hexane, and 3.6 g NdFeB powder (44 wt%). This mixture was homogenized in a Thinky mixer (ARE-250) for 3 min at 2000 rpm. Next, 1 g of this mixture was poured into a Teflon petri dish (3x3 cm, 0.25 mm thickness) and spread evenly by moving a permanent magnet (NdFeB, www.supermagnete.ch) back and forth underneath the petri dish, touching its bottom. Typically, the magnet had a dimension of 5 x 5 x 2 cm and a strength of 0.4 T. We could make different lengths and shapes of pillars by varying the strength of the magnet and the amount of mixture used. Once, pillars had formed, we increased the spacing between the magnet and the petri dish by 0.5 cm and moved this to an oven to be cured for 20 minutes at 60 °C, which solidified the Ecoflex 00-20.

*Size and density of the spheres.* The spherical balls we used for dry SMC cargo transport experiments were as follows. For the experiments in Figure 2, we used 9.5, 8.72, 9.96 mm balls with densities of 1.24 g/cm$^3$ (polycarbonate), 3.83 (alumina) g/cm$^3$, 6.44 (zirconia) g/cm$^3$.

*Glycerol labeling*: To label our samples we added 30 mg of rhodamine isothiocyanate (RITC) dye or 30 mg of green food coloring (Dr Oetker) to 1 ml glycerol. Small aliquots of this were dropped into the native glycerol to visualize the motion induced by our SMCs.

**References and Notes**

1. Bruot, N. & Cicuta, P. Realizing the Physics of Motile Cilia Synchronization with Driven Colloids. *Annu. Rev. Condens. Matter Phys.* **7**, 323–348 (2016).

2. Fauci, L. J. & Dillon, R. Biofluidmechanics of Reproduction. *Annu. Rev. Fluid Mech.* **38**, 371–394 (2006).

3. Gundupalli, S. P., Hait, S. & Thakur, A. A review on automated sorting of source-separated municipal solid waste for recycling. *Waste Manag.* **60**, 56–74 (2017).

4. Majidi, C. Soft-Matter Engineering for Soft Robotics. *Adv. Mater. Technol.* **4**, 1800477 (2019).

5. Alapan, Y. *et al.* Soft erythrocyte-based bacterial microswimmers for cargo delivery. *Sci. Robot.* **3**, (2018).

6. Lee, C.-Y., Chang, C.-L., Wang, Y.-N. & Fu, L.-M. Microfluidic Mixing: A Review. *Int. J. Mol. Sci.* **12**, 3263–3287 (2011).

7. Schaffner, M. *et al.* 3D printing of robotic soft actuators with programmable bioinspired architectures. *Nat. Commun.* **9**, 1–9 (2018).

8. Hu, W., Lum, G. Z., Mastrangeli, M. & Sitti, M. Small-scale soft-bodied robot with multimodal locomotion. *Nature* **554**, 81–85 (2018).

9.  Wehner, M. *et al.* An integrated design and fabrication strategy for entirely soft, autonomous robots. *Nature* **536**, 451–455 (2016).

10. Carpenter, J. A., Eberle, T. B., Schuerle, S., Rafsanjani, A. & Studart, A. R. Facile Manufacturing Route for Magneto-Responsive Soft Actuators. *Adv. Intell. Syst.* **n/a**, 2000283.

11. Rus, D. & Tolley, M. T. Design, fabrication and control of soft robots. *Nature* **521**, 467–475 (2015).

12. Robertson, M. A. & Paik, J. New soft robots really suck: Vacuum-powered systems empower diverse capabilities. *Sci. Robot.* **2**, (2017).

13. Preston, D. J. *et al.* A soft ring oscillator. *Sci. Robot.* **4**, (2019).

14. Vasios, N., Gross, A. J., Soifer, S., Overvelde, J. T. B. & Bertoldi, K. Harnessing Viscous Flow to Simplify the Actuation of Fluidic Soft Robots. *Soft Robot.* **7**, 1–9 (2019).

15. Marchese, A. D., Onal, C. D. & Rus, D. Autonomous Soft Robotic Fish Capable of Escape Maneuvers Using Fluidic Elastomer Actuators. *Soft Robot.* **1**, 75–87 (2014).

16. Kim, Y., Yuk, H., Zhao, R., Chester, S. A. & Zhao, X. Printing ferromagnetic domains for untethered fast-transforming soft materials. *Nature* **558**, 274–279 (2018).

17. Alapan, Y., Yigit, B., Beker, O., Demirörs, A. F. & Sitti, M. Shape-encoded dynamic assembly of mobile micromachines. *Nat. Mater.* **18**, 1244–1251 (2019).

18. Demirörs, A. F., Akan, M. T., Poloni, E. & Studart, A. R. Active cargo transport with Janus colloidal shuttles using electric and magnetic fields. *Soft Matter* **14**, 4741–4749 (2018).

19. Boyraz, P., Runge, G. & Raatz, A. An Overview of Novel Actuators for Soft Robotics. *Actuators* **7**, 48 (2018).

20. Hines, L., Petersen, K., Lum, G. Z. & Sitti, M. Soft Actuators for Small-Scale Robotics. *Adv. Mater.* **29**, 1603483 (2017).

21. Chung, H.-J., Parsons, A. M. & Zheng, L. Magnetically Controlled Soft Robotics Utilizing Elastomers and Gels in Actuation: A Review. *Adv. Intell. Syst.* **3**, 2000186 (2021).

22. Mao, G. *et al.* Soft electromagnetic actuators. *Sci. Adv.* **6**, eabc0251 (2020).

23. Wang, X. *et al.* Untethered and ultrafast soft-bodied robots. *Commun. Mater.* **1**, 1–10 (2020).

24. Bryan, M. T., Martin, E. L., Pac, A., Gilbert, A. D. & Ogrin, F. Y. Metachronal waves in magnetic micro-robotic paddles for artificial cilia. *Commun. Mater.* **2**, 1–7 (2021).

25. Erb, R. M., Martin, J. J., Soheilian, R., Pan, C. & Barber, J. R. Actuating Soft Matter with Magnetic Torque. *Adv. Funct. Mater.* **26**, 3859–3880 (2016).

26. Brennen, C. & Winet, H. Fluid Mechanics of Propulsion by Cilia and Flagella. *Annu. Rev. Fluid Mech.* **9**, 339–398 (1977).

27. Funfak, A. *et al.* Paramecium swimming and ciliary beating patterns: a study on four RNA interference mutations. *Integr. Biol.* **7**, 90–100 (2015).

28. Ramirez-San Juan, G. R. *et al.* Multi-scale spatial heterogeneity enhances particle clearance in airway ciliary arrays. *Nat. Phys.* **16**, 958–964 (2020).

29. Faubel, R., Westendorf, C., Bodenschatz, E. & Eichele, G. Cilia-based flow network in the brain ventricles. *Science* **353**, 176–178 (2016).

30. Gu, H. *et al.* Magnetic cilia carpets with programmable metachronal waves. *Nat. Commun.* **11**, 2637 (2020).

31. Zhang, S., Cui, Z., Wang, Y. & Toonder, J. M. J. den. Metachronal actuation of microscopic magnetic artificial cilia generates strong microfluidic pumping. *Lab. Chip* **20**, 3569–3581 (2020).

32. Dong, X. *et al.* Bioinspired cilia arrays with programmable nonreciprocal motion and metachronal coordination. *Sci. Adv.* **6**, eabc9323 (2020).

33. Milana, E. *et al.* Metachronal patterns in artificial cilia for low Reynolds number fluid propulsion. *Sci. Adv.* **6**, eabd2508 (2020).

34. Hanasoge, S., Hesketh, P. J. & Alexeev, A. Metachronal Actuation of Microscale Magnetic Artificial Cilia. *ACS Appl. Mater. Interfaces* **12**, 46963–46971 (2020).

35. Zhang, S., Cui, Z., Wang, Y. & den Toonder, J. Metachronal μ-Cilia for On-Chip Integrated Pumps and Climbing Robots. *ACS Appl. Mater. Interfaces* **13**, 20845–20857 (2021).

36. Zhang, S., Zhang, R., Wang, Y., Onck, P. R. & den Toonder, J. M. J. Controlled Multidirectional Particle Transportation by Magnetic Artificial Cilia. *ACS Nano* **14**, 10313–10323 (2020).

37. Gauger, E. M., Downton, M. T. & Stark, H. Fluid transport at low Reynolds number with magnetically actuated artificial cilia. *Eur. Phys. J. E* **28**, 231–242 (2009).

38. Jiang, S. *et al.* Three-Dimensional Multifunctional Magnetically Responsive Liquid Manipulator Fabricated by Femtosecond Laser Writing and Soft Transfer. *Nano Lett.* **20**, 7519–7529 (2020).

39. Lu, H. *et al.* A bioinspired multilegged soft millirobot that functions in both dry and wet conditions. *Nat. Commun.* **9**, 3944 (2018).

40. Timonen, J. V. I. *et al.* A Facile Template-Free Approach to Magnetodriven, Multifunctional Artificial Cilia. *ACS Appl. Mater. Interfaces* **2**, 2226–2230 (2010).

41. Rosensweig, R. E. *Ferrohydrodynamics*. (Courier Dover Publications, 1997).

42. Khaderi, S. N., Toonder, J. M. J. den & Onck, P. R. Microfluidic propulsion by the metachronal beating of magnetic artificial cilia: a numerical analysis. *J. Fluid Mech.* **688**, 44–65 (2011).

43. Alapan, Y., Bozuyuk, U., Erkoc, P., Karacakol, A. C. & Sitti, M. Multifunctional surface microrollers for targeted cargo delivery in physiological blood flow. *Sci. Robot.* **5**, (2020).

44. Hussong, J., Breugem, W.-P. & Westerweel, J. A continuum model for flow induced by metachronal coordination between beating cilia. *J. Fluid Mech.* **684**, 137–162 (2011).

45. Gauger, E. & Stark, H. Numerical study of a microscopic artificial swimmer. *Phys. Rev. E* **74**, 021907 (2006).

46. Chateau, S., Favier, J., Poncet, S. & D'Ortona, U. Why antiplectic metachronal cilia waves are optimal to transport bronchial mucus. *Phys. Rev. E* **100**, 042405 (2019).

47. van Leeuwen, J., Aerts, P. & Lowe, C. P. Dynamics of filaments: modelling the dynamics of driven microfilaments. *Philos. Trans. R. Soc. Lond. B. Biol. Sci.* **358**, 1543–1550 (2003).

48. Mathijssen, A. J. T. M., Doostmohammadi, A., Yeomans, J. M. & Shendruk, T. N. Hydrodynamics of micro-swimmers in films. *J. Fluid Mech.* **806**, 35–70 (2016).

49. Vilfan, A. & Jülicher, F. Hydrodynamic Flow Patterns and Synchronization of Beating Cilia. *Phys. Rev. Lett.* **96**, 058102 (2006).

50. Dreyfus, R. *et al.* Microscopic artificial swimmers. *Nature* **437**, 862–865 (2005).

51. Purcell, E. M. Life at low Reynolds number. *Am. J. Phys.* **45**, 3–11 (1977).

**Acknowledgments:** We are much thankful to Prof. André Studart, ETH Zurich for the support and discussions. We also thank the cleanroom facility FIRST at ETH Zurich for instrumental support.

**Funding:**
This research was supported by the Swiss National Science Foundation through the National Centre of Competence in Research Bio-Inspired Materials. J.d.G. thanks NWO for funding through Start-Up Grant 740.018.013 and through association with the EU-FET project NANOPHLOW (766972) within Horizon 2020. A.J.T.M.M. acknowledges funding from the United States Department of Agriculture (USDA-NIFA AFRI Grants No. 2020-67017-30776 and 2020-67015-32330)

**Author contributions:**
A.F.D. proposed and designed the research. D. Z. inspired the research. A.F.D., S.A., S.G., Y.M., E.P., J.C., C.Ü, and D.Z. performed the experiments and analyzed the data. R.H., J.dG., and A.J.T.M.M. performed and analyzed the fluid dynamics simulations. A.F.D., J.dG., and A.J.T.M.M wrote the paper. All authors discussed the results and edited or commented on the manuscript.

**Competing interests:** Authors declare that they have no competing interests.

**Data and materials availability:** All data are available in the main text or the supplementary materials.

Supplementary Material for the paper "**Amphibious Transport of Fluids and Solids by Soft Magnetic Carpets** "

by Ahmet F. Demirörs[*,1], Sümeyye Aykut[1], Sophia Ganzeboom[1], Yuki Meier[1], Robert Hardeman[2], Joost de Graaf[2], Arnold J. T. M. Mathijssen[3], Erik Poloni[1], Julia A. Carpenter[1], Caner Ünlü[4], and Daniel Zenhäusern[5]
E-mail: ahmet.demiroers@mat.ethz.ch

**Fabrication of soft magnetic pillars with various aspect ratios**

Soft magnetic pillars with various aspect ratios were prepared for two sets of samples with different particle concentrations. For both sets the amount of Ecoflex was 4 g (2 g of each component) while the amount of NdFeB particles was changed from 3.6 g (47.4 wt%) to 6.6 g (62.3 wt%). All components were weighted and mixed in a cup and filled into a Teflon dish (area = 4.5x4.5 cm$^2$, thickness = 0.10 mm). To achieve different aspect ratios, the poured amount was varied from 0.5 to 2.5 g. The mixture was evenly spread by moving a permanent magnet (area = 5x5 cm$^2$, thickness = 2 cm, strength = 0.4 T, N pointing upwards, Supermagnete AG) and moving it back and forth and to the sides. Once the mixture was homogeneously distributed, the magnet was moved in circular motions to grow pillars. It is best to start from a distance of about 15 cm below the Teflon dish and to continue the circular motion while slowly moving upwards. After the pillars were formed, the magnet was placed underneath the sample with a spacing of 0.5 cm and cured in the oven at 60 °C for 20 min to solidify the Ecoflex.

**Analysis of the length and aspect ratio of the pillars**

The aspect ratio of the pillars was analyzed using an optical Keyence Microscope. With the 3D stitching software of the microscope, images containing about 40 pillars were recorded. 10 pillars were then chosen and their length and diameter were measured and averaged. These values are tabulated in Table S1.

**Table S1. The length and aspect ratio of the pillars.** The length of the magnetic pillars ranges from about 1000 μm up to 5 mm. The table provides the aspect ratio of the two sample sets with different magnetic-powder content. The aspect ratio decreases with increased thickness of the poured layer (exception: 0.5 g of 3.6 NdFeB).

| 3.6 g NdFeB | Amount filled in PTFE dish [g] | Height h [μm] | Radius r [μm] | Aspect ratio (h/2r) |
|---|---|---|---|---|
| | 0.5 | 944.1 | 139.8 | 3.4 |
| | 1.5 | 2141.8 | 277.1 | 3.9 |
| | 2 | 3356.1 | 585.7 | 2.9 |
| | 2.5 | 3910.6 | 803.2 | 2.5 |
| 6.6 g NdFeB | 0.5 | 3254.2 | 360.7 | 4.6 |
| | 1.5 | 3149.2 | 377.7 | 4.2 |
| | 2 | 5225.9 | 669.6 | 3.9 |

**Model for Solid Cargo Transport**

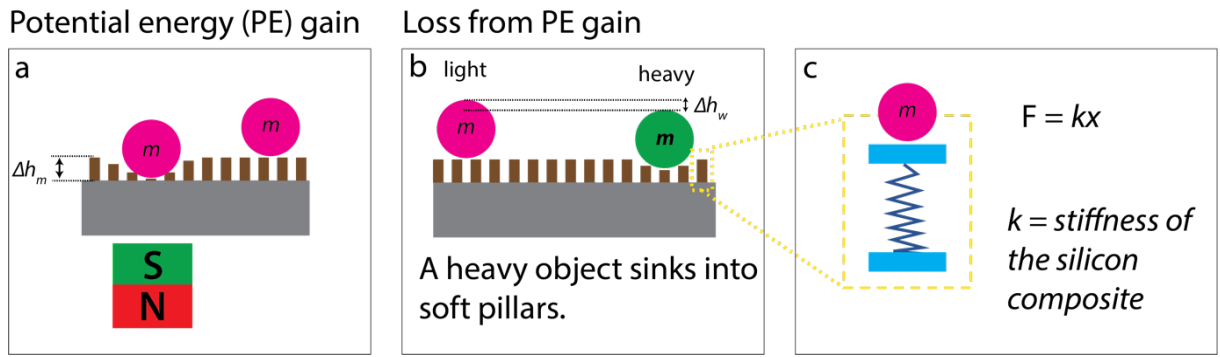

**Figure S1. Minimal model for solid cargo transport on a SMC.** a) The potential energy gain of a sphere sitting in a magnetic well can be estimated from the height difference $\Delta h_m$ between the sphere inside the well and if it were resting on unperturbed pillars. b) A solid sphere will partially sink into the soft pillars, due to its mass, which improves the stability of the well. c) In our model we treat each pillar as a spring that obeys Hooke's law.

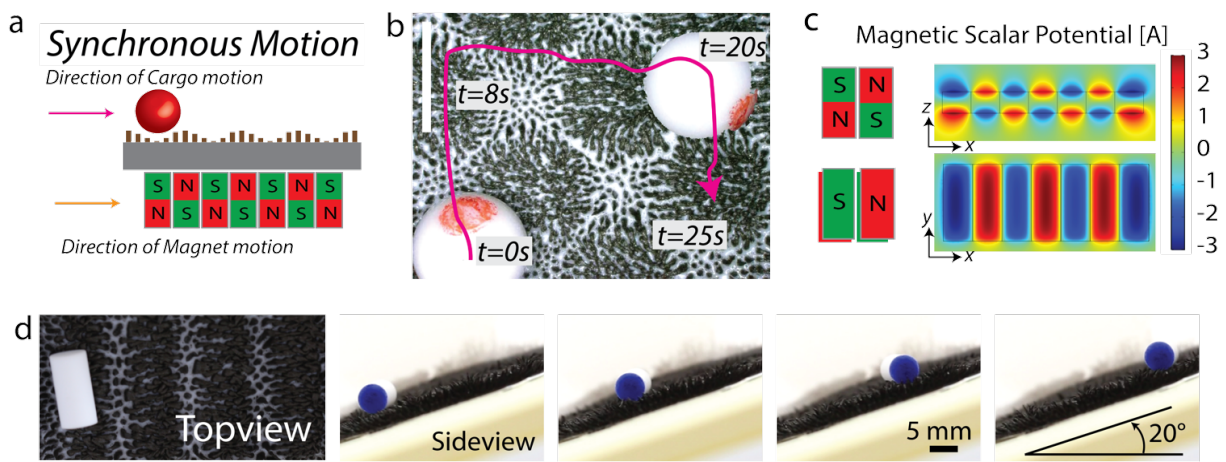

**Figure S2. Solid Cargo Transport with SMCs over a Magnet set.** a) Sketch of a SMC carrying a spherical cargo synchronized with the motion of the magnet. b) Microscopy images taken at different time intervals showing the initial and final configuration of the sphere (the travelled path is indicated in pink), see also Supplementary Movie S3. c) Finite-element calculations and associated sketches of the magnet arrays used in (d) and Figure 4a (main text), showing the side view and top view of the magnetic potential. d) Transport of a cylinder over a ramp at an angle of 20° with respect to the horizontal. Scale bar indicates 5 mm.

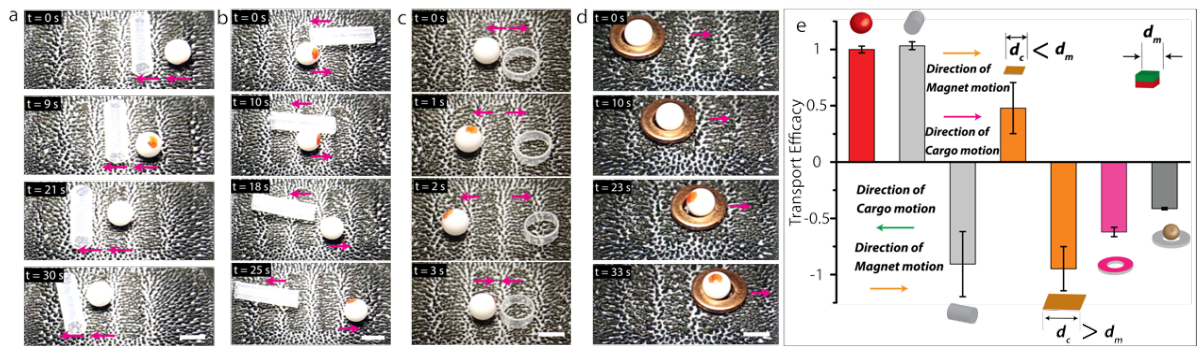

**Figure S3. Solid cargo transport efficiency and particle separation.** a) Two types of solid cargo, a rod and a sphere, are being transported in adjacent wells in the same direction. b) When the rod-shaped cargo is orthogonal to the magnetic wells it is transported in the direction opposite to the magnet and that of the sphere. This demonstrates a rudimentary particle separation mechanism. c) Oppositely directed transport of two types of cargos: a ring and a sphere. d) When a sphere is inserted inside a ring, a combined transport in the direction to that of the magnet occurs. That is, the crowd surfing ring dominates the motion. This coupling lowers the efficacy of the transport. e) An overview of all types of cargo shapes and their transport efficacy with respect to their synchronization with the magnetic field wave. Scale bars indicate 500 μm.

Here, transport efficacy is defined as the ratio between the motion of the object and that of the magnetic track resulting in a number between -1 and 1. When a cargo completely co-moves with the magnetic field wave, the efficacy is at its maximum value of 1. For a cargo that moves completely opposite to the motion of the magnetic field, the efficacy is -1. Different types of solid cargos were tested for their transport direction and efficacy, see Figure S3. The fact that the cargo particles move either same or opposite to the magnet motion allows us a platform to separate particles by moving them in opposite directions. Examples of such separations are demonstrated in Figure S3b-c.

**Sperm number estimation**

We estimated the Sperm number using the expression $S_p = L / \left( \frac{\kappa}{\xi_\perp \, \omega} \right)^{1/4}$ where $L$ is the length of the pillar, $\kappa$ is its bending rigidity, $\omega$ the angular frequency of the magnetic field, and $\xi_\perp$ is the perpendicular drag coefficient. We consulted the paper of Dreyfus *et al.*[50] for these estimations. Sperm number depends on the viscosity of the liquid media and the aspect ratio of the pillars. The estimated Sperm numbers for various viscosities and various aspect ratio pillars are tabulated in Table S2.

**Table S2. Viscosity and Sperm numbers**. Viscosity of the liquid media and the estimated Sperm numbers for the various aspect ratio pillars in such media. Blue labeled viscosity highlights the stall-flow behavior.

|  | Aspect ratio → | 2.4 | 2.9 | 3.4 | 4 |
|---|---|---|---|---|---|
|  | Viscosity | Sperm Number for NdFeB 3.6 g | | | |
| Water | 8.90E-04 | 1.8 | 2.1 | 2.5 | 2.9 |
| 1% PVP | 4.10E-03 | 2.6 | 3.1 | 3.6 | 4.2 |
| 2% PVP | 8.40E-03 | 3.1 | 3.7 | 4.4 | 5.1 |
| 3% PVP | 1.51E-02 | 3.6 | 4.3 | 5.1 | 5.9 |
| 4% PVP | 2.17E-02 | 4 | 4.7 | 5.5 | 6.4 |
| Glycerol | 1.41E+00 | 11.3 | 13.3 | 15.7 | 18.3 |

**The Computational Fluid Dynamics Study**

Our model is comprised of a single sphere with unit diameter (we denote the radius with $r$) that is subject to four forces, which causes it to move along a trajectory that is representative of the cilium dynamics in the experimental system. The model is represented in Figure S4, which shows the three forces and torque acting on it. Here, $\boldsymbol{F}_d$ represents the periodic driving, hydrodynamic interaction with its surrounding leads to the hydrodynamic force $\boldsymbol{F}_h$, the elastic extension is captured by a non-linear spring force $\boldsymbol{F}_s$, and bending of the cilium is accounted for using a non-linear angular torque $\boldsymbol{T}$ that ensures that the bead cannot approach the surface too closely. We will explain our choices for the various forces next.

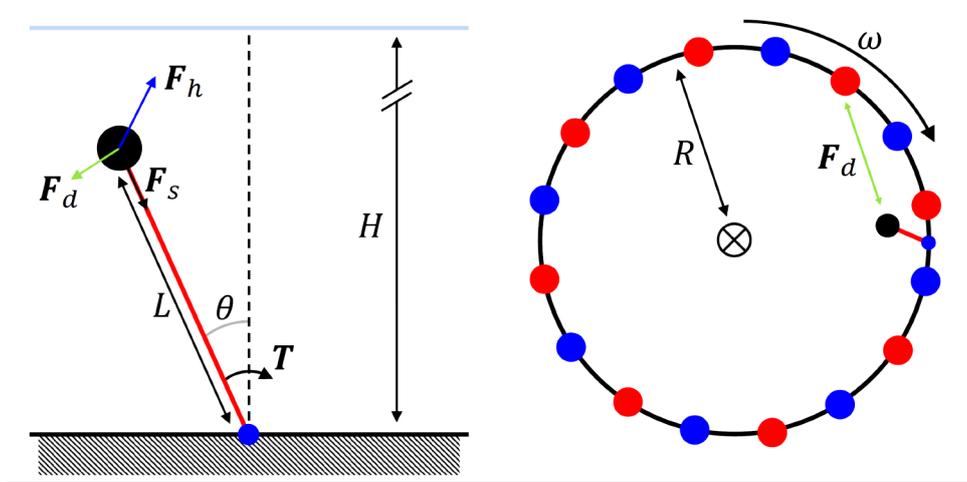

**Figure S4**: **Cartoon of the model cilium**. (left) Cartoon of the single-bead (black disk) model for the cilium, anchored (blue disk) to a no-slip wall (hashing) and bounded from above by a no-shear interface (light blue). A nonlinear angular torque $\boldsymbol{T}$ and spring force $\boldsymbol{F}_s$ (black arrows) cause the bead to move in a roughly semicircular path. The torque depends on the angle $\theta$ with respect to the $z$-axis. The bead is actuated using a periodic driving force $\boldsymbol{F}_d$ (green arrow) and it experiences a hydrodynamic drag due to the presence of a confined fluid medium $\boldsymbol{F}_h$ (blue arrow). (right) Cartoon representing the origin of the periodic driving force. Sixteen points, eight repulsive (blue) and eight attractive (red) push and pull on the bead, respectively, with cubically decaying interaction strength. These points move along a circular track with radius $R$ and angular velocity $\omega$ leading to a net motion of the fluid layer directly above the cilium.

## The Hydrodynamic Boundary Conditions and Mobility

We are interested in the dynamics of a cilium in a thin fluid layer. To model this, we consider a set of hydrodynamic boundary conditions representative of this situation. We place a no-slip surface at $z = 0$, i.e., a zero fluid velocity $\boldsymbol{u}(z = 0) = \boldsymbol{0}$, which represents the bottom surface of our setup. To account for the liquid-air interface, we impose a no-shear boundary condition at $z = H$. This requires that the derivative of the fluid velocity in the direction normal to the interface is zero at the interface. We determine the flow field and mobility of the particles in this system using a Greens' function approach that satisfies Stokes' equations, following Ref.[42].

There is no closed-form expression for the hydrodynamic monopole moment for this thin-fluid-layer geometry. We instead utilize a truncated series of hydrodynamic images using the standard Blake tensor image convention to approximatively account for the no-slip surface and a reflection to account for the no-shear surface. Here, we truncate the series to second order, such that the fluid velocity boundary condition at the interface is satisfied; a more significant truncation error is incurred at the bottom no-slip surface. The fluid velocity $\boldsymbol{u}$ is given by

$$\boldsymbol{u}(\boldsymbol{r}_f + \boldsymbol{r}_s) = \underline{\boldsymbol{g}}(\boldsymbol{r}_f, \boldsymbol{r}_s)\boldsymbol{F}_t \,,$$

where $\boldsymbol{r}_f$ is a point in the fluid, $\boldsymbol{r}_s$ is the point where the force is applied, $\underline{\boldsymbol{g}}(\boldsymbol{r}_f, \boldsymbol{r}_s)$ is the approximate Greens' function, and $\boldsymbol{F}_t$ is the total force acting on the point $\boldsymbol{r}_s$. Figure S5 shows the resulting flow field due to a mid-channel point force oriented along and perpendicular to the surface, respectively. Note that the streamlines follow the top surface, as required by the no-shear boundary condition.

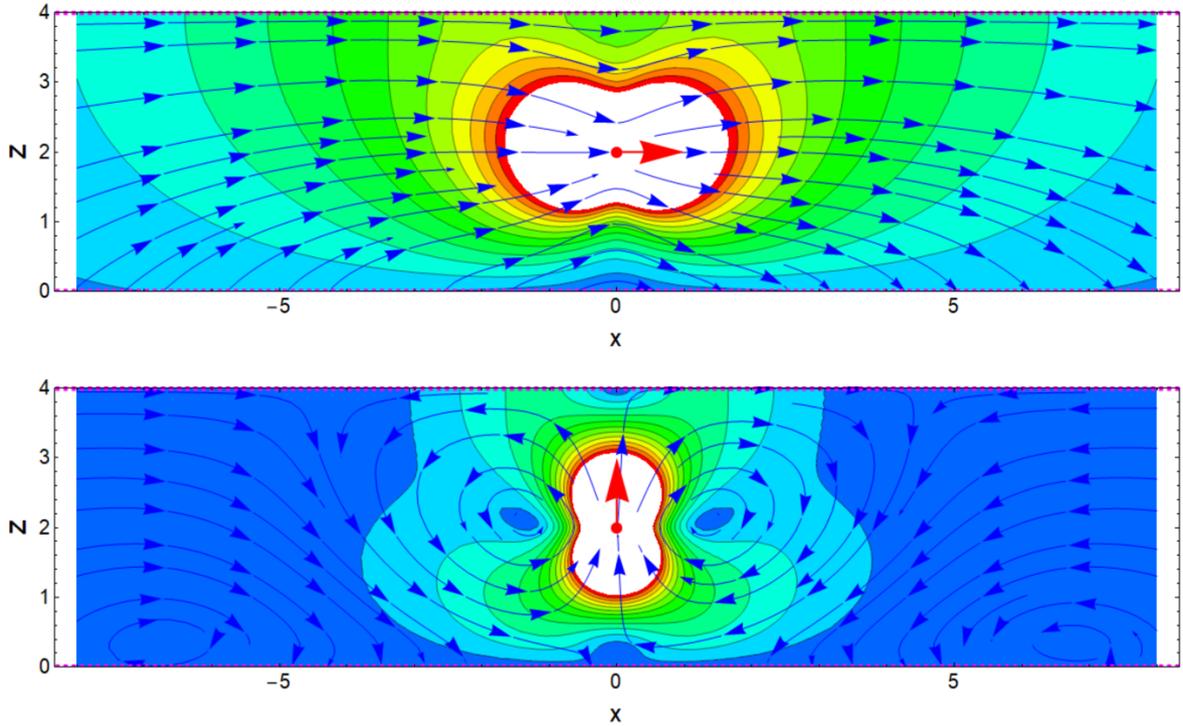

**Figure S5**: **Flow fields generated by the cilium**. Two-dimensional $xz$-slice through the three-dimensional flow field generated by a point force confined between a solid wall with no-slip boundary condition ($z = 0$) and a no-shear fluid-gas interface ($z = 4$); these are indicated using magenta dashed lines. The flow fields have been computed up to the second order to ensure that the boundary condition at the liquid-air interface is well respected. The top panel shows the flow field around a mid-plane force directed along the $x$-axis (red arrow), whilst the bottom panel shows the same situation with the force directed along the $z$-axis. The blue arrows

indicate streamlines, while the color gradient indicates the fluid velocity magnitude, which is increasing from blue to red. In the white region, the flow speed exceeds our cut-off for visualization.

We derived the self-interaction and pair-interaction Rotne-Prager-Yamagawa-type mobility tensors from our monopole flow field Greens' function using the method outlined in Ref. [37]. The process of converting this into a functional and efficient hydrodynamic code is explained in the files and notes enclosed in our accompanying open-access data package. The Mathematica notebook therein may be used to derive higher-order corrections to the Greens' function. However, the compilation and calculation time do not scale favorably with increasing numbers of reflections.

The dynamics of a single bead may be written as

$$\dot{r}(t) = \underline{\mu}\big(r(t)\big)F_t(t)\,,$$

where $\dot{r}(t)$ is the time derivative of the beads position vector (as a function of time $t$), $\underline{\mu}\big(r(t)\big)$ is the self-mobility tensor that depends on the position of the bead within the channel, and which coverts the total force $F_t$ into a velocity. It should be noted that the source code provided with this paper provides additional mobility tensors for pair interactions between beads.

<u>Linear and Angular Spring</u>:

We capture the stretchability and bending of the cilium for our single-bead model using a linear and angular spring, respectively. We ensure that the deviation of the tip is not too significant (the stretching of the cilium is limited) and the tip does not scrape the no-slip surface by imposing finite-extensible non-linear elastic (FENE) responses. We should note that we also explored a version of the multi-bead approach of Refs. [37,45], to be reported on in future work. However, this model proved to be unstable for the significant bending observed in experiment, due to the choice of a dot-product-based angular potential between the beads. In the experimental system, the tip of the cilium scrapes the bottom surface, which is a hydrodynamically challenging situation to resolve and which can strongly limit the numerical stability of any algorithm. These considerations prompted us to utilize the following FENE form for the torque:

$$\boldsymbol{T} = K_a \frac{\theta}{1 - (\theta/\theta_0)^2}\hat{\boldsymbol{e}}$$

in our modeling. Here, $\theta$ is the angle between the cilium and the normal vector to the plane, $\theta_0 = 1.25 \approx 0.4\pi$ is the maximum permitted deviation (sufficiently small to keep the single bead away from the wall), $K_a$ the bending prefactor, and $\hat{\boldsymbol{e}}$ the unit vector that can be constructed from the cross product of the orientation of the cilium ($\hat{\boldsymbol{l}}$ being the unit vector pointing from the anchoring point to the position of the cilium-representing bead) and the normal of the plane $\hat{\boldsymbol{z}}$, i.e., $\hat{\boldsymbol{e}} = \hat{\boldsymbol{l}} \times \hat{\boldsymbol{z}}$. We convert the torque into a force acting on the bead using the lever principle. That is, the torque can be represented as a force $\boldsymbol{F}_a$ acting on the tip of an arm equal in length to the distance between the anchoring point and the cilium bead. The direction is then determined using the cross product of $\hat{\boldsymbol{e}}$ and $\hat{\boldsymbol{l}}$.

The expression for the linear FENE force on the bead, which keeps it from moving too far away from the anchoring point, is given by

$$\boldsymbol{F}_s = K_l \frac{\Delta L}{1 - (\Delta L / \Delta L_0)^2} \hat{\boldsymbol{l}},$$

where $K_l$ is the spring constant, $\Delta L$ is the deviation from the spring's equilibrium length, and $\Delta L_0$ is the maximum permitted deviation from the equilibrium length. We chose $K_l = K_a$ and $\Delta L_0$ to be 5% of $\Delta L_0$ in our calculations. This ensures that our bead followed a closed, area-enclosing path through phase space, necessary for the generation of a net fluid flow[45]. Typically, our model cilium is elongated during the forward stroke, whilst being compressed during the back stroke. The effectiveness of the stroke can be increased by increasing $\Delta L_0$. The speed of the two parts of the stroke is different, as we will also see shortly.

The Periodic Forcing:

To model the force of the magnets acting on the cilium, we account for a periodic driving. Here, we apply the form

$$\boldsymbol{F}_d = K_m \sum_{i=1}^{16} \text{sign}(i) \left( \frac{2r}{|\boldsymbol{p}_i|} \right)^3 \hat{\boldsymbol{p}}_i,$$

where $K_m$ is the magnetic strength, and $\boldsymbol{p}_i$ the position difference between the cilium bead and the $i$-th out of 16-point forces that model the 8 magnets:

$$\boldsymbol{p}_i = \boldsymbol{r}(t) - \begin{pmatrix} R\sin(2\pi i/16 + \omega t) \\ R\cos(2\pi i/16 + \omega t) \\ 0 \end{pmatrix}$$

for which $\omega$ represents angular velocity of the magnetic track and $t$ is the time. We model the decay of the magnetic strength using an inverse cubic scaling here, though other choices can be made. It should be noted that this model does not accurately capture the exact nature of the magnetic field or the motion of the cilium carpet. However, it shows that the features observed in the experiment are robust to significant changes, which is satisfying from the perspective of implementing variants of our soft magnetic carpets. That is, the motion reversal will presumably persist.

Results:

Together, the above considerations give rise to the following expression for the net force acting on a single bead $\boldsymbol{F}_t = \boldsymbol{F}_a + \boldsymbol{F}_s + \boldsymbol{F}_d$, for which the steady-state dynamics is shown in Figure S6. This figure visualizes the motion of the cilium in the low-viscosity co-flowing state and the high-viscosity counter-flowing state. We converted the viscosity $\eta$ in our calculations to an effective Sperm number

$$Sp^* = L \left( \frac{(6\pi \eta a)\omega}{K_a} \right)^{1/4},$$

where $L$ is the equilibrium length, $a$ is the radius of the bead and $\omega = 2\pi/T$. This is a slightly different expression from that used in describing the experiment in the main text. However, this is justified as it only relies on parameters that are present in the simulation; we do not have a rod-like mobility, as is the case in the experiment.

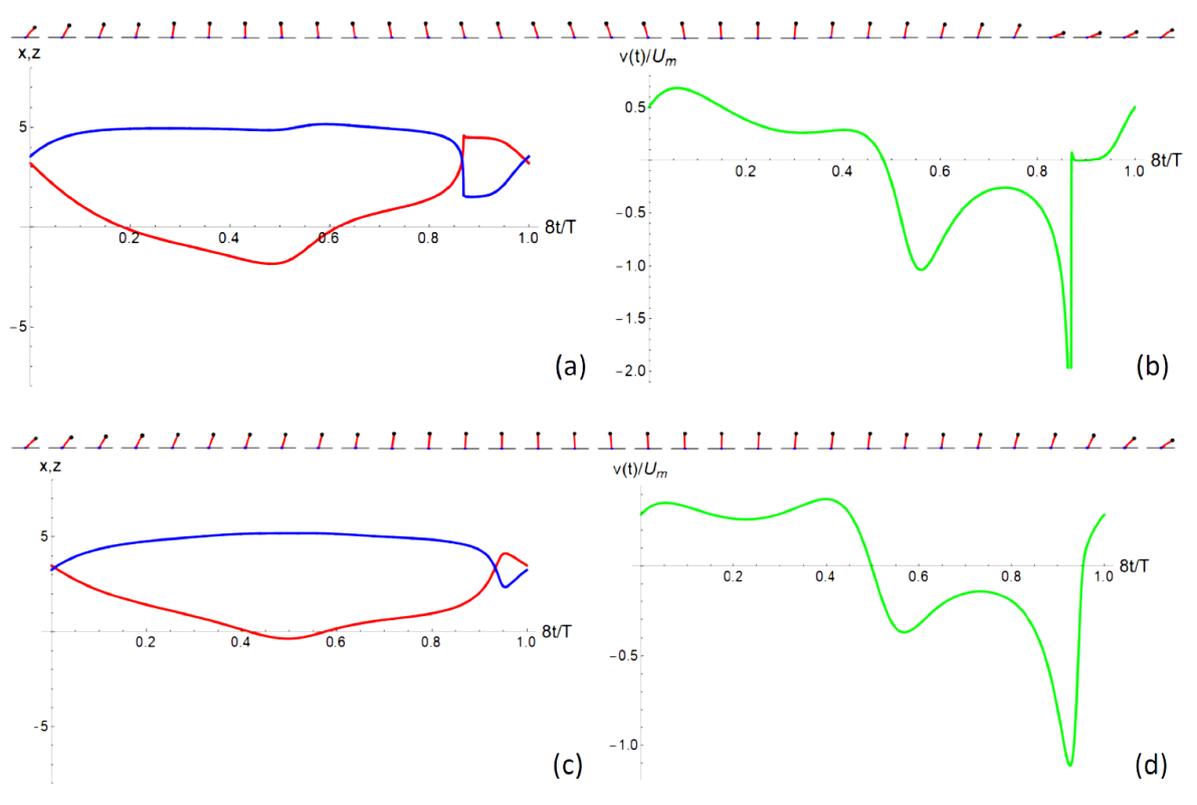

**Figure S6**: **Dynamics of the cilium at different Sperm numbers**. Visualization of the dynamics of our bead-spring cilium model for a fluid layer of height $H = 2L$, with $L$ the cilium length and two sperm numbers $Sp^* = 1.0$ (a,b) and $Sp^* = 1.2$ (c,d), corresponding to the situations of forward (low viscosity regime) and backward flow (high viscosity regime), respectively. Note that positive $v(t)$ corresponds to motion in the direction of the model magnets. Subfigures (a,c) show the position of the bead in the $xz$-plane as a function of the time it takes one of the eight 'magnets' (an attractor-repulsor pair) to pass underneath the cilium. The red curve shows the $x$-coordinate, whilst the blue one shows the $z$-coordinate. Subfigures (b,d) show the instantaneous velocity $v(t)$ of the fluid interface directly above where the cilium is anchored as a function of the same reduced time as in (a,c). The velocity has been reduced by that of the magnets $U_m$. The two strips above the sets of panels show 32 snapshots of the cilium taken equidistantly over the period of motion; forward motion is toward the right. The thin black line indicates the surface, the blue dot the position of the bead, and the red line connects the bead to its anchoring point.

From Figure S6 it becomes clear that within our simple model, the transition from co-fluid motion to counter-fluid motion is due the interplay between the driving and elastic properties of the cilium. The reversed motion relies on the cilium achieving its maximum angular extent for some time, as the model magnetic track runs underneath it. The rapid motion toward the surface is indicative of the attractive part of the 'magnetic' field moving in unison with the bead. This causes the bead to be pushed close (or the tip of the pillar in the experiment even onto) the surface. The reduced mobility close to the surface leads to there being a retention time, as only the repulsive 'magnetic' interaction is sufficient to push the bead away. This retention eliminates the most powerful part of the return stroke, thus giving rise to net forward motion of the interface just above the anchoring point. In the case of a greater viscosity, the motion toward the surface cannot be made fully, leading to less significant effects of near-wall mobility reduction. That is, the period of slow-down that is visible in Figure S6b is nearly absent in Figure S6d. Here, the forward motion loses out against the backward motion, as the return stroke may be entered in a lower-mobility regime (closer to the wall), but it the force closer to the wall is more substantial. The secondary (weak) motion reversal, not shown here, is caused by the linear extensibility of the cilia dominating the physics of our model at even higher viscosities, *i.e.*, when the deflection is minimal. We believe this to be predominantly a result of our choices of modeling the magnetic track, which we will explore in future work. Lastly, it should be noted that it is the net fluid motion of the interface above the cilium that reverses, rather than the motion between cilia, as can occur when multiple cilia are modelled[28].

**Supplementary Movies**

Movie S1. Circular transport of a sphere on dry soft magnetic carpet.
Movie S2. Transport of a sphere on a rectangular path on dry soft magnetic carpet.
Movie S3. Cycle of single cilium motion with the travelling magnetic field.
Movie S4. Linear transport of a board-shaped cargo on dry soft magnetic carpet.
Movie S5. Particle separation on dry soft magnetic carpet.
Movie S6. Uphill linear transport of a cylinder on dry soft magnetic carpet.
Movie S7. Linear Transport of a glycerol droplet on soft magnetic carpet.
Movie S8. Flow generated by the wet soft magnetic carpet in glycerol.
Movie S9. Spatially controlled liquid mixing with wet soft magnetic carpet in viscous medium.
Movie S10. Comoving liquid flow with wet soft magnetic carpet in DI water.
Movie S11. Stall-flow of the liquid on the wet soft magnetic carpet in an intermediate viscosity medium.
Movie S12. Oppositely-directed liquid flow generated with the wet soft magnetic carpet in a viscous medium.